\documentclass[12pt,a4paper]{article}
\usepackage{amsmath,amssymb,bm,graphicx}
\usepackage[CP866]{inputenc}
\usepackage[english,russian]{babel}
\makeatletter
\renewcommand\section{\@startsection {section}{1}{\z@}%
                                   {-3.5ex \@plus -1ex \@minus -.2ex}%
                                   {2.3ex \@plus.2ex}%
                                   {\normalfont\Large\bfseries\centering}}
\makeatother

\newcommand{\sr}{\selectlanguage{russian}}
\newcommand{\se}{\selectlanguage{english}}
\newcommand{\karti}[5][]{\begin{figure}[#2]\begin{center}%
\includegraphics[height=#3]{#4.ps}
\end{center}#1\caption{#5}\end{figure}}
\newcommand{\ssy}[5]{\emph{#1}. #2. #4.\ V.~#3.\ #5\rlap{.}}
\newcounter{primer}
\newcommand*{\parnum}[2][]{\refstepcounter{primer}\par\smallskip\noindent
\textbf{#2 \theprimer.#1}}
\DeclareMathOperator{\Bd}{Bd}
\DeclareMathOperator{\supp}{supp}

\newcommand{\rea}{ I\!R}
\newcommand{\mink}{I\!L}
\newcommand{\sph}{\ensuremath{\mathbb S}}
\newcommand{\evkl}{I\!E}
\newcommand*{\ogr}[2]{#1\,\vrule \,{}_{\displaystyle{}_{#2}}}
\newcommand{\tot}{\ensuremath{E_{\rm tot}^-}}
\newcommand*{\rmd}{\mathrm{d}}
\title{СВЕРХСВЕТОВЫЕ ПЕРЕМЕЩЕНИЯ В (ПОЛУ)КЛАССИЧЕСКОЙ ОТО}
\date{}
\author{С. В. Красников\thanks{Главная астрономическая обсерватория РАН}}
\begin{document}
\maketitle
\begin{abstract}
Рассматривается возможность сверхсветовых перемещений \glqq
не\-тахионных\grqq{} (т.~е.\ движущихся с мгновенной скоростью $v<c$) тел в
рамках общей теории относительности. Обсуждается запрет, предположительно
налагаемый на такие перемещения квантовой теорией поля. Демонстрируется
несостоятельность этого запрета в общем случае.
\\[10mm]
    Ключевые слова: сверхсветовые перемещения, причинность, кротовина,
    квантовое неравенство.
\end{abstract}
\section{СВЕРХСВЕТОВЫЕ ПЕРЕМЕЩЕНИЯ}

\flqq Скорость  тела, состоящего
из обычной (\glqq нетахионной\grqq) материи не может превышать скорость
света\frqq. Если  под \glqq скоростью\grqq\  понимать \glqq мгновенную
скорость\grqq , т.~е.\ скорость в точке, то это утверждение \emph{верно}, но
\emph{тавтологично} --- если найдется
тело, летящее быстрее света, его придется назвать тахионом. На практике,
однако, под скоростью в данном случае понимается нечто иное (нечто вроде
эффективной средней скорости, как мы увидим). Утверждение оказывается
\emph{содержательным} и
\emph{неверным}.

Считается, что расстояние между Землей и  Денебом составляет $d=1500$ световых
лет. Допустим, корабль сумел долететь от Земли до Денеба за время
$T_1<1500\,$лет, или --- ниже мы увидим, что это совершенно разные допущения
--- сумел долететь до Денеба и вернуться на Землю
 за $T_2<3000\,$лет (в обоих случаях время измеряется по земным часам).
Представляется вполне законным назвать такое перемещение \glqq
сверхсветовым\grqq{}, поскольку $d/T_1>c$. Как мы увидим чуть ниже на
конкретных примерах, сверхсветовое --- в этом смысле --- перемещение возможно,
в принципе, даже для тела, движущегося с мгновенной скоростью $v<c$. Это
связано с природой величины $d$: в общем случае в ОТО едва ли можно придать
разумный смысл понятию \glqq расстояния между телами\grqq{}, а когда это
оказывается возможным, расстояние вовсе не обязано совпадать с тем, что выше
было обозначено $d$. Утверждение \flqq расстояние между Землей и Денебом
составляет $1500\,$световых лет\frqq{} следует, строго говоря, понимать
следующим образом. Есть область вселенной $N$, включающий в себя часть мировой
линии Земли, часть мировой линии Денеба и некоторые из соединяющих их
изотропных геодезических. Область эта (приблизительно) плоская и мы можем
рассматривать ее и как часть реального пространства-времени (с его, возможно,
очень сложной геометрией), и как часть
\emph{фиктивного} пространства Минковского, в котором мировым
линиям Земли и Денеба соответствуют параллельные прямые. Понятие расстояния
между такими прямыми вводится очевидным образом. Именно это расстояние и
составляет $1500\,$световых лет.

Как явствует из вышесказанного, сверхсветовые перемещения удобней обсуждать не
в терминах скоростей, а сравнивая реальное про\-стран\-ство-время с
пространством Минковского \cite{portal}.
\paragraph*{Опеределение.} Пусть $C$ --- времениподобный цилиндр
$\sum_{i=1}^3x_i^2\leqslant r_0^2$ в пространстве Минковского  \mink$^4$.
Область $U$ глобально гиперболического пространства $M$, а иногда и само $M$,
будем называть \emph{лазом} (соответствующий английский термин --- shortcut),
если существует изометрия $\varkappa\colon\; (M-U)\to(\text{\mink}^4-C)$  и
пара точек $p,q\in (M-U)$ таких, что
$$
 p\preccurlyeq q,\qquad\varkappa(p)\not\preccurlyeq\varkappa(q)
$$
\par\noindent
(напомню обозначения: $a\preccurlyeq b$ означает, что из точки $a$ до $b$
можно провести направленную в будущее причинную кривую; ниже нам еще
понадобятся $J^+(a)\equiv\{x|\:a\preccurlyeq x\}$,
$J^-(a)\equiv\{x|\:x\preccurlyeq a\}$ и $J^\pm(\gamma)\equiv
\bigcup_{a\in\gamma} J^\pm(a)$). Итак, $M$
--- лаз, если его можно получить из  пространства Минковского заменой цилиндра
$C\subset
\text{\mink}^4$ на нечто такое (а именно, на $U$), что точки,
пространственноподобно разделенные в \mink$^4$, становятся причинно
связанными. Путешествие сквозь лаз из $p$ в $q$ как раз и есть  сверхсветовое
перемещение, т.~к. тело попадает туда, куда свет (будь это пространство
Минковского) дойти бы не успел.

\parnum[ \glqq Пузырь Алькубиерре\grqq .]{Пример}\label{ex:Alc}
На плоскости \rea$^2$ рассмотрим метрику
$$
\rmd s^2 = \Omega^2(r)(\rmd x^2 +\rmd y^2) ,\qquad r\equiv\sqrt{x^2+y^2}
$$
где  $\Omega$ --- монотонная функция такая, что
$$
\ogr{\Omega}{r>r_0 }=1,\quad
\ogr{\Omega}{r<(1-\delta)r_0}=\Omega_0=const,\qquad
0<\delta,\Omega_0\ll 1 .
$$
Все пространство, кроме колечка $1-\delta<r/r_0<1$  является плоским и, более
того, при $r>r_0 $ оно неотличимо от Евклидовой плоскости. Тем не менее, точка
$(x=-r_0 $, $y=0)$ намного ближе ($\approx 2\Omega_0 r_0 $ против $2r_0 $) к
диаметрально противоположной точке $(x=r_0 ,$ $y=0)$, чем в случае Евклидовой
плоскости. Естественное 4-мерное обобщение (ср.~\cite{Alc})
$$
\rmd s^2 = -\rmd t^2 + \Omega^2(r)(\rmd x_1^2 + \rmd x_2^2+ \rmd x_3^2)
\qquad r\equiv\Big(\sum_{i=0}^3x_i^2\Big)^{1/2}
$$
очевидно является лазом [см.~рис.~\ref{fig:lazy}(а)].
\karti[\hfil (a)\hspace{14 em} (б)\hfil]{h}{9.8 em}{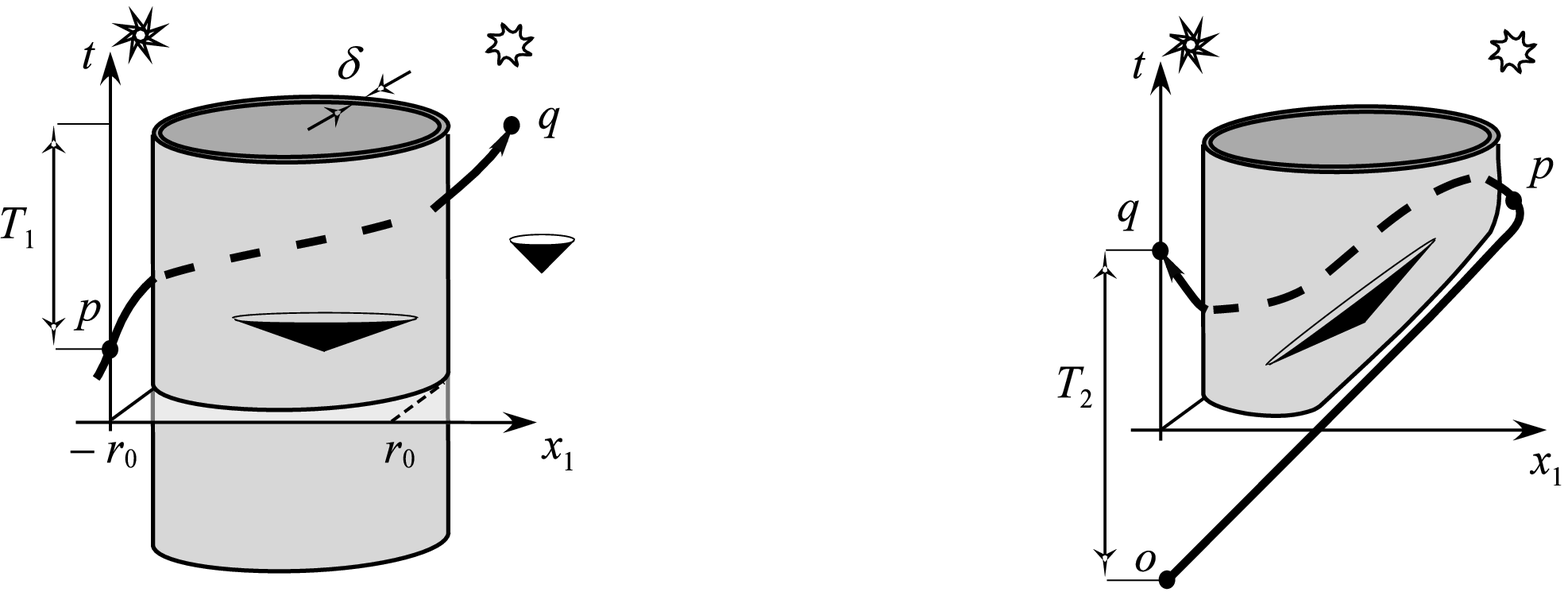}{(a)
Световой конус внутри лаза раскрыт (в данных координатах) в $1/\Omega_0$ раз
шире, чем в пространстве Минковского. Поэтому пунктирная линия времениподобна.
(б) \glqq Искусственный\grqq{} лаз. $o$ предшествует ему.
\label{fig:lazy}}
\parnum[ \glqq Труба Красникова\grqq .]{Пример}\label{ex:tube} Пусть $M$
 --- это \rea$^4$ с метрикой \cite{FTL}
\begin{equation}\label{eq:tr}
\rmd s^2 = (\rmd x_1-\rmd t)(k\rmd x_1+\rmd t) + \rmd x_2^2+ \rmd x_3^2,
\end{equation}
где $k=k(r)$ --- монотонная функция такая, что
$$
\ogr{k}{r>r_0}=1,\quad
\ogr{k}{r<(1-\delta)r_0}=k_0=const,\qquad
-1<k_0<0 .
$$
Пространство, как и в предыдущем примере, искривлено только в сферическом слое
толщины $\delta$. Главное
--- и, как мы увидим ниже, важное --- отличие состоит в том, что путешествие теперь может закончиться
\emph{еще до того как началось}, см.~рис.~\ref{fig:lazy}(б). Эффект этот чисто
координатный (время измеряется в разных точках) и с нарушением причинности не
связан. Действительно, рассмотрим функцию $F=t+\frac{k_0-1}{2}x_1$ и любую
непродолжимую причинную кривую $\lambda(\zeta)$, где $\zeta$ --- натуральный
параметр во вспомогательной евклидовой метрике
\[
\rmd s_E^2 = \omega_0^2+\omega_1^2 + \rmd x_2^2+ \rmd x_3^2,
\qquad \omega_0\equiv \rmd t +\tfrac{k-1}{2}\rmd x_1,
\quad \omega_1\equiv \tfrac{k+1}{2}\rmd x_1
\]
(она отличается от метрики \eqref{eq:tr} только знаком при $\omega_0^2$).
Нетрудно доказать, что $|\rmd F/\rmd\zeta|>(1+k_0)/2$, откуда ясно, что
$\lambda$ должна рано или поздно пересечь (единственный раз) поверхность
$F=0$. Последняя, т.~о., является поверхностью Коши $M$ и, следовательно (см.
\cite[предложение~6.6.3]{HawEl}), $M$  глобально гиперболично. Наличие
требуемых $p$ и $q$ очевидно, и мы заключаем,   что $M$ ---  лаз.

\parnum[ \glqq Кротовина\grqq .]{Пример}\label{ex:wh} Вырежем из пространства
Минковского два тонких цилиндра, см.~рис.~\ref{fig:krot}(а), склеим границы
получившихся дыр и
\glqq сгладим\grqq{} метрику в окрестности склейки  так, чтобы
устранить возникший там разрыв производных.
\karti[\hfil (a)\hspace{20 em} (б)\hfil]{h}{14 em}{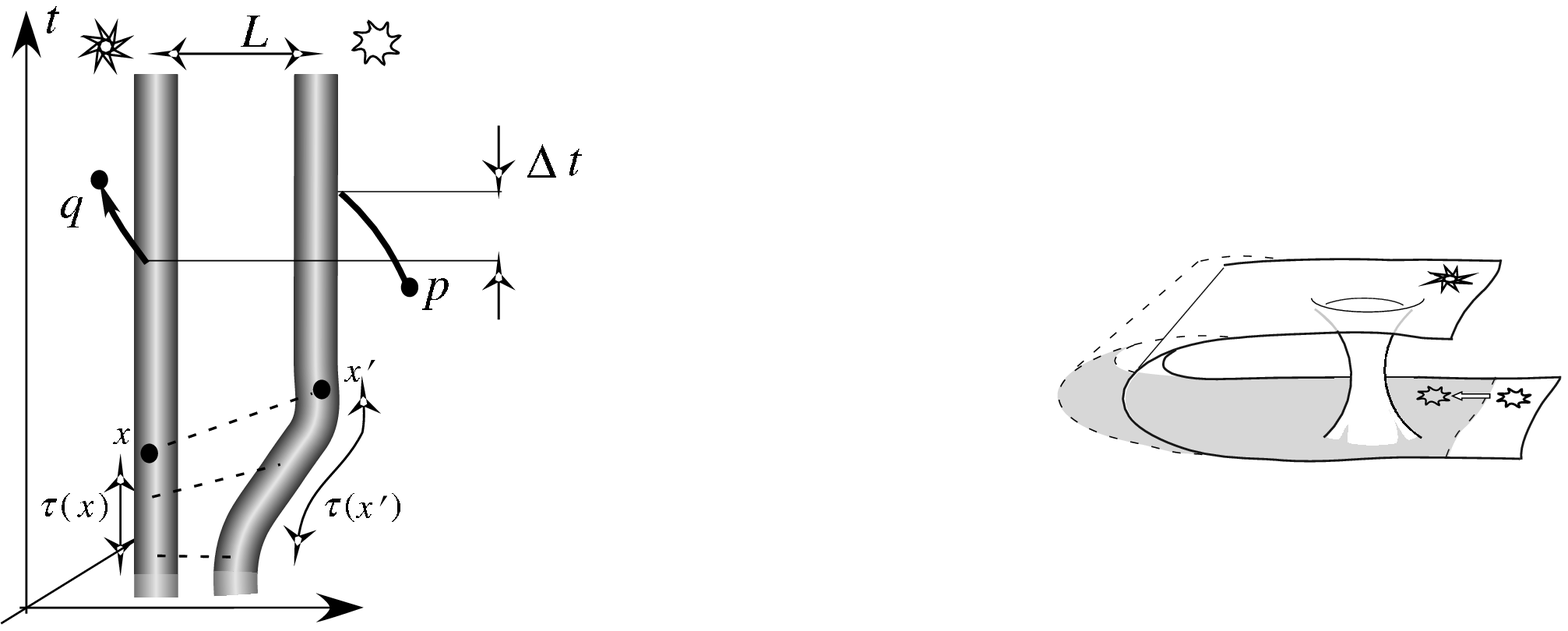}{Кротовина с
меняющимся во времени расстоянием между входами.\label{fig:krot} Концы
пунктирных линий склеены.} Получившееся пространство \cite{Tho,viss}, называют
\emph{кротовиной} (в англоязычной литературе --- \se  wormhole\sr{}). Часто
так же называют и его пространственноподобное сечение, рис.~\ref{fig:krot}(б).
Описанная процедура неоднозначна, т.~к. точки на границах цилиндров можно
отождествлять по-разному. Обычно склеивают точки с равным $\tau$, где $\tau$
есть (лоренцево) расстояние до некоторой плоскости --- скажем, $t=0$ ---
измеренное вдоль образующей цилиндра. Такое правило нужно, если мы хотим
описывать кротовину, эволюционирующую, как показано на рис.~\ref{fig:krot}(б):
расстояние между входами во \glqq внешнем\grqq{} пространстве меняется, а
форма кротовины (и, соответственно, ее \glqq длина\grqq{}) --- нет. Очевидно,
если при $t>0$ один вход покоится, а другой движется, то у склеиваемых точек
окажется разное $t$ (\glqq парадокс близнецов\grqq{}). Поэтому, кротовину
можно (в частности) использовать и для перемещений, которые \glqq кончаются
раньше, чем начались\grqq{}, см. ~рис.~\ref{fig:krot}(а). Как и в прошлом
примере, это не обязательно нарушает причинность: если $\Delta t < L$,
пространство остается глобально гиперболичным и, как следствие, --- лазом.

Пригодность лазов для межзвездных перелетов различна. Будем говорить, что
точка $o\in M$ \emph{предшествует} лазу, если $\varkappa$ может быть
продолжена на $M- J^+(o)$, и  назовем лаз
\emph{вечным}, если ему не предшествует никакая точка.
Очевидно, в отсутствие тахионов лаз можно интерпретировать как созданный по
решению, принятому в $p$, только если  $p$ этому лазу предшествует. Вечный же
лаз создать вообще нельзя, а можно только
\emph{найти}.

Все три рассмотренных выше лаза, будучи статическими, являются, конечно,
вечными. В случае кротовины это обстоятельство принципиально --- в глобально
гиперболическом про\-стран\-стве-вре\-мени кротовина не может ни появиться, ни
исчезнуть (см. \cite[предложение~6.6.8]{HawEl}). В то же время, два первых
лаза легко модифицировать так, чтобы они перестали быть вечными. Достаточно
отменить  при $t,r<r_0$ условия наложенные  на $\Omega$ и $k$,  а при $t<r$
потребовать, чтобы $\Omega$ и $k$ были равны 1. Таким лазам предшествует,
например, начало координат.

К сожалению,  точка $p$, фигурирующая в определении лаза, не может этому лазу
предшествовать (иначе $q$ --- в противоречии с определением
--- лежала бы в $M- J^+(p)$, поскольку $\varkappa(q)$ лежит в \mink$^4 -
J^+[\varkappa(p)]$). Т.~о., когда решение лететь к Денебу принято, строить лаз
уже поздно в том смысле, что $T_1$ это не уменьшит. Что, однако, не
обесценивает идею такого строительства --- лазом еще не поздно воспользоваться
для уменьшения $T_2$ (см.~рис.~\ref{fig:lazy}(б)), а это, похоже, для
практических нужд гораздо важнее.

\section{КВАНТОВЫЕ ОГРАНИЧЕНИЯ}
До сих пор мы обсуждали возможность преодоления \glqq светового барьера\grqq{}
с чисто геометрической точки зрения. Однако, в рамках ОТО геометрия вселенной
и свойства заполняющего ее вещества связаны уравнениями Эйнштейна $G_{ik}=8\pi
T_{ik}$ (здесь и ниже по умолчанию подразумевается система единиц, в которой
$G =\hbar = c = 1$), и можно поставить вопрос о том, какими свойствами должен
обладать тензор энергии-импульса (ТЭИ) материи, заполняющей лаз. Оказывается,
во всех рассмотренных выше примерах есть точки, где $G_{ik}t^i t^k<0$ для
некоторого времениподобного вектора $\bm t$ и, значит, нарушается т.~н.\
слабое энергетическое условие (СЭУ), требующее, чтобы
$$
T_{ik}t^i t^k\geqslant 0\qquad\quad \forall \,\bm t\colon \ t^i t_i\leqslant
0,
$$
т.~е. чтобы плотность энергии была  неотрицательна в любой системе отсчета.
Хотя строгое доказательство пока найдено только для кротовин
\cite{FSW}, похоже, что нарушение СЭУ присуще
\emph{любому} лазу. Грубо говоря, если бы плотность энергии всюду была
неотрицательна, то естественно было бы ожидать, что на больших расстояниях
(т.~е.\ в Ньютоновском пределе) гравитационное поле убывало бы, как $1/r^2$, а
не исчезало совсем, как того требует определение лаза.

В классической физике плотность энергии предположительно не может быть
отрицательной (поэтому вещество, нарушающее СЭУ называют \emph{экзотическим})
и т.~о.\ для сверхсветовых перелетов остаются две возможности:

1). Можно исследовать ситуации, в которых нарушения СЭУ связаны просто с тем,
что мы приписали нулевую кривизну области $N$. Тем самым мы
\emph{заранее} посчитали отсутствующими любые источники энергии и, в том числе,
нужные для обеспечения (соблюдающего СЭУ) сверхсветового перелета. Чтобы
последовательно учесть наличие потенциальных источников энергии (скажем,
звезд, лежащих чуть в стороне от прямой Земля--Денеб), понятие лаза придется
обобщить (это не трудно) так, чтобы $M$ сравнивалось не с
\mink$^4$, а с неким --- тоже искривленным --- $M'$. Такого рода лазы,
действительно, не требуют экзотического вещества (см.~Приложение в
\cite{portal}), но вопрос о том, насколько эффективными в этом случае они могут
быть, остается открытым.

2). Можно также учесть квантовые поправки к уравнениям Эйнштейна. В
полуклассической гравитации \cite{GMM,BirDav} последние приобретают вид
\begin{equation*}
G_{ ik }=8\pi T^{\rm C}_{ ik } + 8\pi \, T_{ ik },
\end{equation*}
где $T^{\rm C}_{ ik }$ --- вклад классической материи, а второй член --- это
ренормированное среднее значение ТЭИ рассматриваемого поля. Известно, что $T_{
ik }$ может нарушать СЭУ (классический пример --- эффект Казимира). Эта,
вторая возможность с точки зрения квантовой теории поля гораздо интересней, на
ней мы и остановимся.

Пусть $\gamma(\tau)$ --- мировая линия свободно падающего наблюдателя и $\tau$
--- его собственное время. Рассмотрим \glqq усредненную с $\chi$\grqq\
плотность энергии $\rho = T_{\hat 0\hat 0}$ (шляпки над индексами тензора
означают, что его компоненты определяются в ортонормированном базисе, с
нулевым ортом $\partial_\tau$)

\[
\rho_\chi(\tau_0)\equiv \int_{-\infty}^{-\infty}
\rho(\tau)\chi(\tau-\tau_0)\,\rmd\tau,
\]
где интеграл берется вдоль $\gamma$, а функция $\chi$ нормирована условием
$$
 \int_{-\infty}^{-\infty} \chi(\tau)\,\rmd\tau=1.
$$
В случае скалярного (с минимальной связью) и электромагнитного полей в
$d$-мерном пространстве Минковского ($d=2,4$)  доказано \cite{F&RoIN}, что для
функций специального вида, а именно $ h(\tau)\equiv\pi^{-1}\Delta
/(\tau^2+\Delta ^2) $, справедливо неравенство
\begin{equation}\label{eq:qinOR}
\rho_h> -\Delta ^{-d}.
\end{equation}
Факт этот часто интерпретируется как математическое выражение некоего
\glqq принципа дополнительности\grqq , связывающего нарушения СЭУ
с  промежутком времени $\Delta $, в течение которого его можно наблюдать:  чем
сильнее нарушение (т.~е.\ чем больше $|\rho|$), тем меньше $\Delta $. Исходя
из идеи, что искривленное пространство является
\glqq почти плоским\grqq , если рассматривается достаточно маленькая область,
Форд и Роман предположили \cite{F&Ro96}, что подобный принцип верен
универсально
--- т.~е.\ для любых квантовых состояний любых свободных\footnote{Что для
взаимодействующих полей он заведомо неверен, видно уже из того, что ему не
подчиняется (даже в плоском пространстве) эффект Казимира, в котором
$\rho(t)=const<0$.} полей в любом пространстве-времени. Конкретно, если
усреднение производится с функцией $f$:
\begin{equation}\label{eq:uslnaf}
f\in C^\infty,\qquad \supp f\in (\tau_1,\tau_2),
\qquad
  \int_{\tau_1}^{\tau_2}[f'(\tau)]^2/f(\tau)\,\rmd\tau\lesssim 1,
\end{equation}
то, предположительно \cite{singed}, должно выполняться неравенство
\begin{equation}\label{eq:rrr}
\rho_f\gtrsim -\mathcal T^{-d}\qquad\forall\,\mathcal T\lesssim\mathcal T_R,
\end{equation}
где
\[
\mathcal T\equiv|\tau_2-\tau_1|,\qquad
\mathcal T_R\equiv
\big(\max |R_{\hat \imath\hat \jmath\hat m\hat n}|\big)^{-1/2}.
\]
Максимум здесь берется по всем компонентам и всем $\tau\in (\tau_1,\tau_2)$.
Физический смысл \glqq квантового неравенства\grqq\  \eqref{eq:rrr}  примерно
тот же, что и \eqref{eq:qinOR}, а условие на $\mathcal T$ должно гарантировать
малость рассматриваемой области (иногда его приходится дополнять неким
топологическим требованием, см. раздел~\ref{subs:neodn}). Выбор функций
\eqref{eq:uslnaf} вызван тем, что именно для таких $f$ неравенство
\eqref{eq:rrr} удалось доказать (при $d=2$ для безмассового скалярного поля в состоянии
конформного вакуума \cite{conf2,Few}).

Квантовое неравенство (КН) позволяет весьма эффектно продемонстрировать \glqq
нефизичность\grqq\  лазов, рассмотренных выше \cite{97}\footnote{На самом
деле, в этих работах рассматривались чуть более сложные метрики.}. Ход
рассуждений приблизительно таков. Пусть $p$ такая точка, что через нее
проходит времениподобный геодезический сегмент $\gamma(\tau)$ длины $\mathcal
T_R$, вдоль которого  $\rho$ отрицательна и примерно постоянна. Предположим,
\begin{equation}\label{eq:noC}
\max |R_{\hat \imath\hat \jmath\hat m\hat n}(p)|\lesssim
 \max |T_{\hat k\hat l}(p)|\approx -\rho(p)
\end{equation}
(во всех трех лазах действительно можно найти $p$, удовлетворяющую
предъявленным требованиям). Применяя
\eqref{eq:rrr} с $\mathcal T=\mathcal T_R$ к $\gamma$, получим
\begin{equation}\label{eq:cep}
|\rho_f| <\mathcal T_R^{-4} = \big(\max |R_{\hat \imath\hat \jmath\hat m\hat
n}|\big)^{2}
\lesssim \rho^2(p)
\end{equation}
или
\begin{equation}\label{eq:Pl_den}
|\rho(p)|\gtrsim \rho_f/\rho(p)\approx 1.
\end{equation}
Т. о. плотность энергии в $p$ должна быть Планковской! В рассмотренных выше
примерах можно представить себе, что  СЭУ нарушается только в тонком ---
скажем, Планковской толщины --- сферическом слое. Однако, если мы хотим
использовать лаз для перемещения макроскопических объектов (с характерным
размером $l$), то естественно ожидать, что диаметр слоя будет $\sim l$ и объем
$V\sim\pi l_\text{Pl}l^2$. Поэтому при $l\sim 1\,$м \glqq суммарное количество
отрицательной энергии\grqq{}, заключенное даже в столь тонком слое будет
чудовищным:
\begin{equation}\label{eq:ogr}
|\tot|=-\rho V \sim \pi\left(\frac{2,2\times 10^{-8}\,\text{кг}}{(1,6\times
10^{-35}\text{м})^3}\right) 1\,\text{м}^2 (1,6\times 10^{-35}\text{м})\approx
3\times 10^{62}\,\text{кг}.
\end{equation}
Очевидно, каков бы ни был (несколько туманный) физический смысл величины
$\tot$, этот результат следует интерпретировать как полную невозможность
создания такого рода лазов даже в очень отдаленном будущем. Сам по себе этот
факт не слишком важен в случае пространств из примеров~\ref{ex:Alc},
\ref{ex:tube}. Действительно,  оба пространства выбиралась
настолько простыми, насколько позволяла задача, для которой они были
придуманы: иллюстрация \emph{принципиальной} совместимости сверхсветовых
перемещений с запретом локального превышения скорости света. Можно
предполагать поэтому, что нежелательные свойства требующихся для них
источников являются следствиями именно этой простоты. Проблема, однако, в том,
что рассуждения, приведшие к
\eqref{eq:ogr}, представляются настолько общими, что кажется, будто они
справедливы для
\emph{любого} лаза. Это означало бы, что квантовое неравенство исключает
сверхсветовые перемещения.

\section{УНИВЕРСАЛЬНОСТЬ ОГРАНИЧЕНИЙ}
Общность рассуждений, обосновывающих \eqref{eq:ogr}, в действительности
не\-сколько обманчива. Существует ряд ситуаций, в которых  КН, даже если оно
справедливо, не требует, тем не менее, справедливости \eqref{eq:ogr}.
Рассмотрим некоторые из них (подробности см.~\cite{portal,noQI}).
\subsection{Нелокальные эффекты}\label{subs:neodn}
Как известно~\cite{BirDav}, у безмассового скалярного поля на цилиндре
$$
\rmd s^2 = -\rmd t^2 +\rmd x^2 \qquad
x=x+L.
$$
есть квантовое состояние, в котором $\rho=-\frac{\pi}{6}L^{-2}$. Такая
плотность энергии очевидно противоречит
\eqref{eq:rrr}, поскольку в данном случае $\mathcal T_R=\infty$. Возможно, это
связано с тем, что на масштабах $\gtrsim L$ цилиндр --- хотя он и плоский ---
все же \glqq недостаточно похож\grqq\  на часть пространства Минковского.
Т.~о. условие на величину $\mathcal T$ в \eqref{eq:rrr}
\emph{необходимо дополнить} еще каким-нибудь требованием, которое бы учитывало
такую возможность. Можно, в частности\footnote{Это обобщение условия, которое
Фьюстер \cite{Few} использует в случае $d=2$. Других приемлемых (т.~е., как
минимум, координатно независимых) вариантов подобного условия пока не
предлагалось.}, ограничить действие КН геодезическими $\gamma$ настолько
короткими, что $J_\gamma\equiv J^+(\gamma)\cap J^-(\gamma)$ имеет топологию
шара (как это имеет место в \mink$^4$). Такое условие, однако, выводит из-под
действия КН наиболее интересную разновидность кротовин. Действительно, пусть
$\lambda$  --- любой времениподобный геодезический отрезок в горловине
статической (для простоты) кротовины.  Можно показать, что при  $\Delta t \to
L$ и
\emph{неизменной геометрии горловины} на  $\lambda$ обязательно  найдутся
точки, соединимые \emph{негомотопной}  $\lambda$ причинной кривой (и значит
$J_\lambda$ уж точно не шар). Т.~о., КН оказывается тем менее ограничительным,
чем
\glqq эффективнее\grqq{} кротовина с точки зрения межзвездных перелетов.

\subsection{Искривление пространства удаленными телами}
Важную роль при выводе \eqref{eq:Pl_den} играло неравенство в
\eqref{eq:noC}, выражающее идею,
что кривизна в $p$  возникает за счет наличия в $p$ экзотического вещества. С
учетом уравнений Эйнштейна обсуждаемое неравенство приобретает вид
\begin{equation}\label{eq:bezVeyl}
\max |R_{\hat \imath\hat\jmath\hat m\hat n}(p)|
\lesssim \max |G_{\hat k\hat l}(p)|
\end{equation}
и \emph{на первый взгляд} должно выполняться всегда  при отсутствии
специальных симметрий или подгонок параметров. Это, однако, не верно. Кривизна
в точке определяется (среди прочего) тензором Вейля:
\[
R_{\hat \imath\hat\jmath\hat m\hat n}=C_{\hat \imath\hat\jmath\hat m\hat n} -
 g_{\hat \imath[\hat n}R_{\hat m]\hat\jmath}
  - g_{\hat\jmath[\hat m}R_{\hat n]\hat \imath}
   -\tfrac{1}{3}g_{\hat \imath[\hat m}g_{\hat n]\hat\jmath}R.
\]
Т.~о. основания ожидать, что \eqref{eq:bezVeyl} справедливо, есть только,
когда
\begin{equation*}
P\equiv\max |C_{\hat \imath\hat\jmath\hat m\hat n}|/ \max |R_{\hat
\imath\hat\jmath}|\lesssim 1,
\end{equation*}
а это, \emph{как правило}, не выполняется. В частности, в
\emph{любой} искривленной, но пустой области (например, в окрестности
 любой звезды) $P=\infty$.

\subsection{Неоднородное распределение материи}
Попробуем воспроизвести вывод \eqref{eq:ogr}, опустив условие постоянства
$\rho(\tau)$. Тогда для оправдания последнего неравенства в
\eqref{eq:cep} придется потребовать $|\rho(p)|\approx
\max_\gamma|\rho|$. С другой стороны, для \eqref{eq:Pl_den} нужно
$ |\rho_f|\gtrsim  |\rho(p)|$. Значит для получения таким способом
\eqref{eq:ogr} существенно, чтобы $|\rho_f|\approx \max_\gamma|\rho|$,
т.~е.
\begin{equation}
|\rho_f|\gtrsim |\overline\rho|,\qquad
\overline\rho\equiv
\mathcal T^{-1}\int_{\tau_1}^{\tau_2}\rho(\tau)\,\rmd\tau
\end{equation}
($\overline{\rho}$, очевидно, есть \emph{обычное} среднее). Вкупе с
\eqref{eq:rrr} это значит, что необходимым условием является
\begin{equation}\label{eq:j,sx}
|\overline\rho|\lesssim \mathcal T^{-d} \qquad\text{при}\quad \mathcal
T\lesssim\mathcal T_R.
\end{equation}

Верно ли неравенство \eqref{eq:rrr} за пределами двумерного
конформно-три\-ви\-аль\-ного случая неизвестно, но вот \eqref{eq:j,sx}
определенно нарушается даже и в этом случае \cite{noQI}.

\parnum{Пример}
Рассмотрим
  двумерное пространство де Ситтера. При
подходящем выборе координат $u$, $v$ (покрывающих \emph{все} пространство)
метрика в области $W=\{\epsilon<\arctg e^{2v}<\arctg
e^{2u}<\tfrac{\pi}{2}-\epsilon\}$ примет вид
\[
\rmd s^2= \alpha^2\sinh^{-2}(u-v)\rmd u\rmd v.
\]
ТЭИ безмассового скалярного поля в состоянии конформного вакуума,
ассоциированного с так записанной метрикой \cite{BirDav}
\[
  T_{vv}= T_{uu}=
-\tfrac{1}{12\pi}, \qquad
  T_{uv}= \tfrac{1}{12\pi}\sinh^{-2}(u-v).
\]
Пусть $\gamma$ --- сегмент времениподобной геодезической, заданный условием
\[
v+u=0,\quad t\in[t_1,t_2],
\qquad t\equiv \tfrac{1}{2}(v-u).
\]
Потребуем, обеспечив этим, в частности, $\gamma\subset W$, чтобы
\begin{equation*}
\tfrac{1}{2}\ln\tan\epsilon<t_1<t_2\ll -1.
\end{equation*}
Плотность энергии, измеренная наблюдателем с мировой линией $\gamma$, будет
\[
\rho=T_{\hat 0\hat 0}=
2(T_{uu}-T_{uv})(\alpha^2\sinh^{-2}(u-v))^{-1}
\approx
 -\frac{\sinh^{2}(u-v)}{6\pi \alpha^2},
\]
откуда
\[
-\overline\rho=\frac{\mathcal T^{-1}}{6\pi}\int^{t_2}_{t_1}
\alpha^{-1}\sinh(u-v)\,\rmd t \approx
\frac{\mathcal T^{-1}}{24\pi}\alpha^{-1} e^{-2t_1}.
\]
В то же время
\[
\mathcal T=-\int^{t_2}_{t_1}\,\alpha\sinh^{-1}2t\,\rmd t
\approx \alpha\int^{2t_2}_{2t_1}e^{s}\rmd s
\approx
\alpha e^{2t_2}.
\]
Таким образом,
\[
\overline\rho=-\tfrac{1}{24\pi}e^{2(t_2-t_1)}{\mathcal T}^{-2},
\qquad\mathcal T\ll \mathcal T_R=\alpha,
\]
что очевидно нарушает (к тому же сколь угодно сильно при достаточно малых
$\epsilon$ и $t_1$) неравенство \eqref{eq:j,sx}.
\subsection{\glqq Экономные\grqq{} лазы}
Даже если плотность энергии экзотического вещества $\rho\sim -1$, отсюда в
общем случае все еще не следует \eqref{eq:ogr}. Дело в том, что $V$, на самом
деле, не обязательно должен быть $\gtrsim l^2$.

\parnum[ \glqq Портал\grqq .]{Пример} Введем две положительные константы, $\rho_0$, и
$\eta_\varepsilon$, связанные требованием $\rho_0\gg\eta^2_\varepsilon$, и две
функции ---  произвольную гладкую четную функцию $\varepsilon(\eta)$  со
свойством $\supp\varepsilon=(-\eta_\varepsilon,\eta_\varepsilon)$
 и $\rho$, определенную равенством
\begin{equation*}
 \rho(\eta,\psi)\equiv \rho_0 - \eta^2\cos 2\psi.
\end{equation*}
Рассмотрим пространство-время
\begin{equation}\label{eq:portmet}
\begin{split}
    \rmd s^2=-\rmd t^2 + 4(\varepsilon^2 + \eta^2)(\rmd \eta^2 +
    \eta^2\rmd\psi^2) + \rho^2\rmd\phi^2,\\
\eta,\rho\geqslant 0,\qquad \phi=\phi+2\pi,\quad \psi=\psi+2\pi.
\end{split}
\end{equation}
(подразумевается, что точки с $\eta=0$, различающиеся только значением $\psi$,
отождествляются, так же как и точки с $\rho=0$, различающиеся только значением
$\phi$). Чтобы увидеть структуру этого пространства-времени (подробнее
см.~\cite{portal}), заметим, что при $\eta>\eta_\varepsilon$ метрика имеет вид
$$
\rmd s^2 =-\rmd t^2 +
 \rmd z^2 + \rmd \rho^2 + \rho^2\rmd\phi^2,
$$
где $z\equiv \eta^2\sin 2\psi $. Поскольку $z(\psi)=z(\pi+\pi)$, можно
представить пространство
\eqref{eq:portmet}, как результат следующей хирургии.
Возьмем два экземпляра Евклидова пространства \evkl$^3$ и вырежем из каждого
по диску $\mathcal D$ радиуса $\rho_0$. Склеим теперь эти разрезанные
пространства: левый берег каждого разреза отождествляется с правым берегом
другого. Получившееся пространство (на самом деле, это просто двулистное
накрытие \evkl$^3 -\sph$, где $\sph=\Bd \mathcal D$) сингулярно, т.~к. мы
вынуждены удалить точки ветвления $\sph$ (в них теряется структура хаусдорфова
многообразия). Оказывается, однако, что если должным образом искривить его
метрику в полнотории (толщины $\eta^2_\varepsilon$ в нашем случае), окружающем
$\sph$, то указанная сингулярность
\emph{устранима}. В пространство можно вклеить окружность так, что оно станет
полно. Это и будет cечение $t=const$ пространства-времени \eqref{eq:portmet}.

Чтобы получить из
\eqref{eq:portmet} лаз, достаточно удалить по полупространству --- скажем,
$z>d$ и $z<-d$ --- из каждого листа, а появившиеся границы склеить. Результат
показан на рис.~\ref{fig:geom}(а).
\karti[\hfil (a)\hspace{14 em}\hspace{4em} (б)\hfil]{h}{14em}{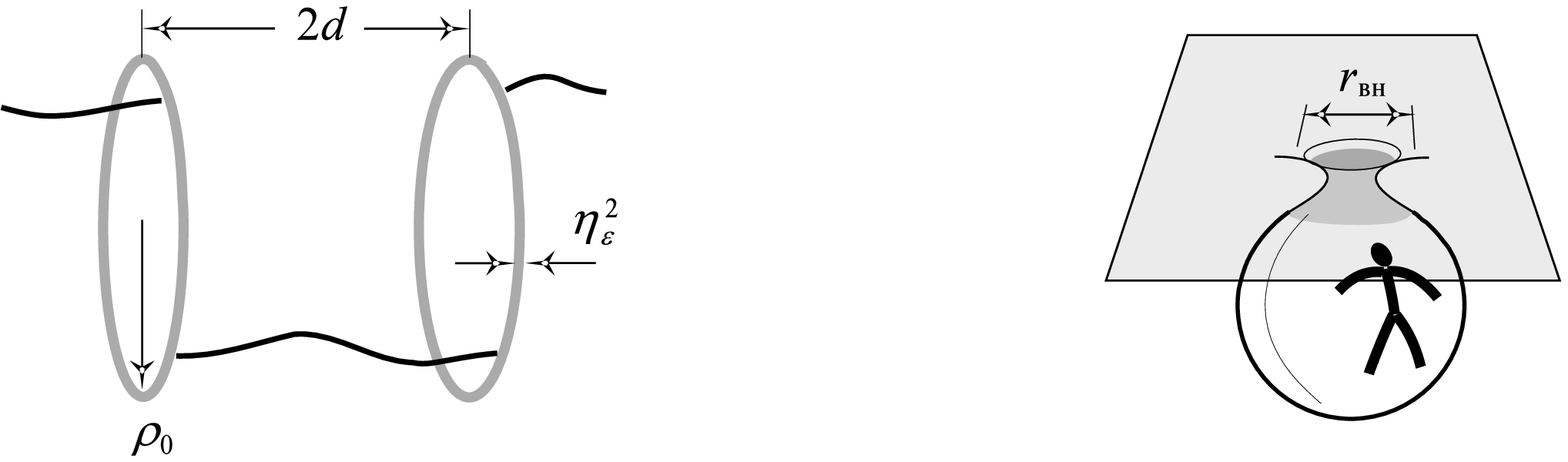}{(а)
\glqq Портал\grqq{}. Два серых кольца, на самом деле, представляют собой
\emph{единый} полноторий. Верхняя толстая кривая непрерывна. Нижняя ---
является нестягиваемой петлей. (б) \glqq Карман Ван Ден Брука\grqq{}. СЭУ
нарушается только в горловине (более темный участок).
\label{fig:geom}}

Макроскопическое тело может пройти сквозь портал с $\rho_0\sim l/2$ при том,
что область ненулевой кривизны в таком портале имеет объем $V\sim\pi l$ (я
выбрал $\eta_\varepsilon\sim 1$). Даже если она вся заполнена экзотическим
веществом планковской плотности, то $|\tot|$ при $l\sim 1\,$м составит
\hbox{$\sim 5\times 10^{27}\,$кг}. Хотя это и очень большая величина, она все же
на 35 порядков (!) меньше той, что предсказана
\eqref{eq:ogr} и составляет \glqq всего лишь\grqq\  $\sim 2\times 10^{-3} M_\odot$.

Следует также отметить, что лаз, пригодный для транспортировки
макроскопических тел не обязан \emph{сам} быть макроскопическим,
обстоятельство позволяющее, в принципе, уменьшить \tot\ еще на 35 порядков.
Дело в том, что даже очень большой объем может быть ограничен сферой очень
маленькой площади. Например, в пространстве, показанном на
рис.~\ref{fig:geom}(б) область заполненная  экзотическим веществом
--- это сферический слой радиуса $r_\text{вн}$. И какою бы большой ни была
область с пассажиром, $r_\text{вн}$ может оставаться микроскопическим
\cite{Broeck}. Можно, в частности потребовать $r_\text{вн}\sim 1$. Для
удержания от схлопывания такого \glqq кармана\grqq{} достаточно, оказывается,
всего $\tot\sim -10^{-3}\,$г экзотической материи \cite{portal}.

\end{document}